\documentclass{moriondNOTITLES}

\def\be{\begin{equation}}
\def\ee{\end{equation}}
\def\bea{\begin{eqnarray}}
\def\eea{\end{eqnarray}}

\def\fsat{f_{\scalebox{0.9}{\rm sat}}}
\def\fgw{f_{\scalebox{0.9}{\rm gw }}}
\def\DaAvg{\langle\Delta a\rangle}

\usepackage[symbol]{footmisc}  

\newcommand{\UCLouvain}{Centre for Cosmology, Particle Physics and Phenomenology - CP3, Université Catholique de Louvain, Louvain-La-Neuve, B-1348, Belgium}
\newcommand{\ROB}{Royal Observatory of Belgium, Avenue Circulaire, 3, 1180 Uccle, Belgium}

\newcommand{\Photo}{}

\begin{document}
\vspace*{4cm}
\title{Probing gravitational waves using GNSS constellations}

\author{Soumen Roy${}^{1,2,}$\:\footnote{\texttt{\href{mailto:soumen.roy@uclouvain.be}{soumen.roy@uclouvain.be}}}, Bruno Bertrand${}^{2}$, and Justin Janquart${}^{1,2}$}
\address{${}^{1}$\UCLouvain \\ ${}^{2}$\ROB}



\maketitle
\abstracts{
The detection of gravitational waves opened up a new window to look into the Universe by probing phenomena invisible through electromagnetic observations. As gravitational waves interact very weakly with matter, their detection is challenging and expensive. So far, they have been observed in the nHz frequency and audible ranges. Future detectors are expected to cover the mHz frequency, leaving the $\mu$Hz regime largely unexplored.
With on-board atomic clocks and orbits determined to the cm, Global Navigation Satellite System constellations (GNSS), like GPS or Galileo, offer free access to more than 30 years of clock and orbit data for tests of general relativity. 
We develop a framework for calculating the deviation in the evolution of GNSS orbits induced by gravitational wave signals. 
We show that when a gravitational wave interacts in resonance with a satellite's orbit, effects amplify, which can be used to bridge the gap in the $\mu$Hz regime. 
 Finally, we demonstrate that the orbital deviations induced by gravitational waves are coherent across the entire constellation, enabling a satellite network to disentangle GW effects from satellite systematics.
}



\section{Introduction}
The direct detection of gravitational waves (GWs) has opened a new window into observational astronomy and fundamental physics~\cite{GW150914}. To date, the LIGO–Virgo–KAGRA detector network has identified several hundred GW events from compact binary mergers, primarily within the frequency band of 10–1000 Hz. More recently, the NANOGrav collaboration reported evidence for a stochastic GW background in the nano-Hz range~\cite{NANOGrav:2023gor}, likely originating from inspiralling supermassive black hole binaries.

To fully exploit the scientific potential of GWs, a number of next-generation detectors are being planned for terrestrial and space-based deployment, extending sensitivity into the mHz to kHz range. However, building new detectors is not only technically challenging but also extremely costly, and a significant observational gap remains in the $\mu\rm{Hz}$ regime. Bridging the full GW spectrum will therefore require innovative and cost-effective detection strategies targeting currently inaccessible frequency bands.

One possibility is to study the interaction of GWs with binary systems. When a GW signal is incident on a binary system, it produces tidal forces that stretch and squeeze the spacetime between the components of the binary. This effect is amplified when the GW frequency is a multiple integer of the binary orbital frequency, which is known as tidal resonance~\cite{Mashhoon1978}. For a long-duration signal, this effect accumulates over time. This secular drift in orbital parameters can be used to reconstruct the signal properties. In this work, we investigate the GW-induced tidal resonance effect on GNSS constellation through a series of numerical simulations. To study the potential offered by a network of satellites, we consider a list of Galileo satellites, with initial positions taken from a snapshot on 2016-11-21 00:00:00 UTC~\footnote{\href{https://www.gsc-europa.eu/system-service-status/orbital-and-technical-parameters}{www.gsc-europa.eu/system-service-status/orbital-and-technical-parameters}}.

\section{Orbit Modelling}
To model the orbit of a satellite, the Earth-satellite system is typically treated as a Keplerian two-body problem governed by Newtonian gravity. The orbit is described using six parameters $(a, e, \iota, \Omega, \omega, \nu)$, known as Keplerian elements, where $a$ is the semi-major axis, $e$ is the orbital eccentricity, $\iota$ is the inclination angle, $\Omega$ is the right ascension of the ascending node, $\omega$ is the argument of periapsis, and $\nu$ is the true anomaly. In the absence of external forces, these elements remain constant over time. However, satellites are always influenced by external forces. To determine how the orbit evolves under such perturbations, we typically employ the Gauss planetary equations~\cite{brouwer1961methods}. These equations describe the time evolution of the osculating elements in response to the applied perturbing accelerations. The osculating elements can be described as the instantaneous ellipse fitted to the satellite's position and velocity $(x, y, z, \dot{x}, \dot{y}, \dot{z})$ with respect to the central mass. We evolve the orbit in this Cartesian representation using Cowell's formulation, as implemented in the \texttt{poli\textcolor{blue}{astro}} package~\cite{juan_luis_cano_rodriguez_2022_6817189}.

As the Earth–satellite system forms a freely falling frame with its center of mass fixed at the origin, the perturbative force on satellites due to an incident GW signal can be derived from the geodesic deviation equation~\cite{MaggioreVol1}:
\begin{equation}
\mathcal{F}_i = \frac{1}{2} \ddot{h}_{ij} x^j,
\end{equation}
where $ x^j \equiv \{x, y, z\} $, and $h_{ij} $ is the transverse-traceless metric perturbation. The metric perturbation is given by \mbox{$h_{ij}(t) = e^+_{ij}(\hat{\mathbf{n}}) h_+(t) + e^\times_{ij}(\hat{\mathbf{n}}) h_\times(t)$}, with polarization tensors $e^+_{ij} = u_i u_j - v_i v_j$ and $e^\times_{ij} = u_i v_j + v_i u_j$. Here, $\hat{\mathbf{n}}$ denotes the GW propagation direction, while $\hat{\mathbf{u}}$ and $\hat{\mathbf{v}}$ are orthonormal unit vectors transverse to $\hat{\mathbf{n}}$. A detailed formulation of the polarization tensors in cylindrical coordinates is provided in Ref.~\cite{Blas:2021mpc}. We then transform into the geocentric Cartesian coordinate system to ensure compatibility with the \texttt{poliastro} framework.



\section{Effect of an incident GW on GNSS satellites}

The perturbing effect of a GW signal on a satellite orbit depends on two factors: phase of the incident GW signal and the Keplerian parameters of the satellite's orbit. The nature of the response depends on whether the satellite’s orbital motion and the incident GW signal are in-phase or out-of-phase. In the out-of-phase case, the orbit shrinks, while it stretches when they are in-phase. Here, we investigate which effects play the most important role in the detectability of the effect and how one can maximize the detection potential using the GNSS system.



\begin{figure}
\centering
\begin{minipage}[b]{0.6\textwidth}
  \includegraphics[width=\linewidth]{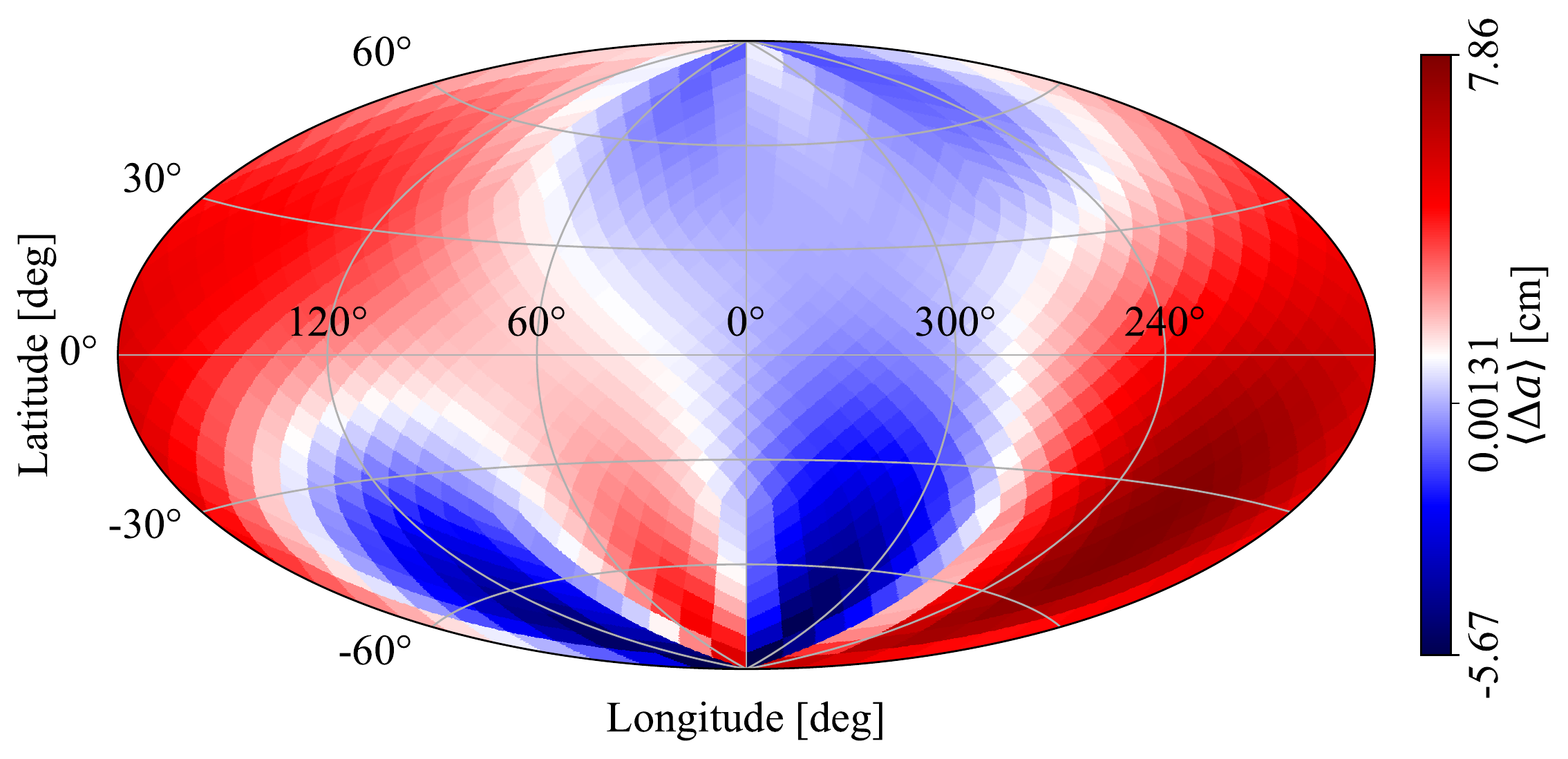}
\end{minipage}%
\hfill
\begin{minipage}[b]{0.38\textwidth}
  \caption{Response to sky position of the GW source for \texttt{GSAT0102}. We show the orbit-averaged secular drift in semi-major axis $\DaAvg$ after an observation period of 100 days with a GW signal at a frequency of $\fgw=3\fsat$ and the strain amplitude $h_0 \sim 10^{-12}$. The dark red indicates the sky patch responsible for maximum orbit expansion, and dark blue corresponds to maximum orbit shrinkage.}
  \label{fig:skymap}
\end{minipage}
\end{figure}

We first investigate how the orbital response varies with the sky position of the GW source, assuming a single monochromatic plane wave with frequency $\fgw=3\fsat$, where $\fsat$ is the satellite's orbital frequency. Figure~\ref{fig:skymap} shows the result for the \texttt{GSAT0102} satellite in a circular orbit. The sensitivity varies significantly across the sky, with an enhancement factor of $|\DaAvg|_{\rm max} / |\DaAvg|_{\rm min} \sim 6\times10^3$, where $\langle \cdot \rangle$ denotes an orbit-averaged quantity. The dark blue region near the south pole is highly sensitive to orbital shrinkage; such high-sensitivity spots generally exist, but they will change place depending on the GW-satellite system one chooses. It is important to note that the sky-dependent sensitivity differs for each harmonic.

\begin{figure*}
\centering
\includegraphics[width=0.97\textwidth ]{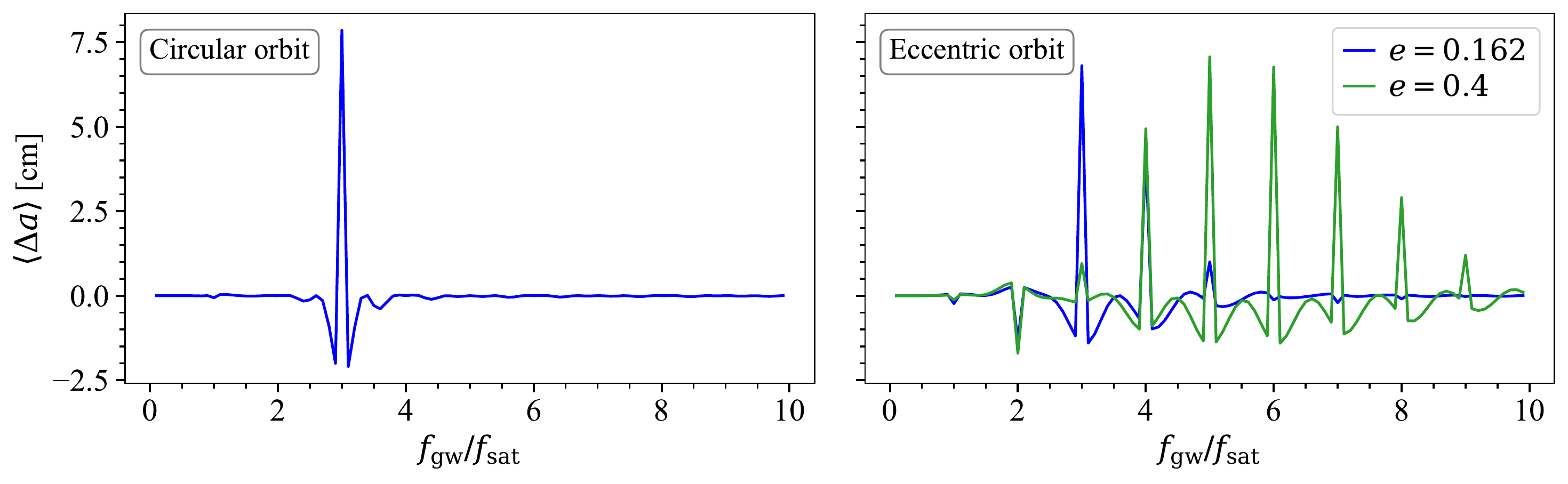}
\caption{Illustration of the orbit averaged secular drift in the semi-major axis $\langle\Delta a\rangle$ of the satellite due to the perturbation of continuous GW signal at frequency of $\fgw$ and the strain amplitude $h_0 \sim 10^{-12}$. We show the results of $\langle\Delta a\rangle$ after an observation period of 100 days as a function of $\fgw /\fsat$. The left and right (blue) panels show the results for \texttt{GSAT0102} and \texttt{GSAT0201}, respectively. The green plot in the right panel is obtained by setting $e=0.4$ for \texttt{GSAT0201}, where the fifth harmonic is most dominant. Tenth becomes most dominant when $e=0.6$.
}
\label{fig:CircEcc}
\end{figure*}

We now explore how the tidal response changes with the GW frequency and orbit eccentricity, showing that higher harmonics become more important for higher eccentricities. We consider \texttt{GSAT0102} with a circular orbit and \texttt{GSAT0201} with an eccentric orbit of $e=0.162$. We evolve the orbits assuming various frequencies for the incident GW, fixing the sky position where $\DaAvg$ is maximum in Figure~\ref{fig:skymap}, and evaluate $\DaAvg$ assuming a 100-day observation period. Figure~\ref{fig:CircEcc} shows $\DaAvg$ as a function of $\fgw/\fsat$. For \texttt{GSAT0102} (left panel), resonances occur at $\fsat$ and $3\fsat$, with the latter clearly dominating, reaching an enhancement factor of $\mathcal{O}(10^3)$. However, when the GW source is positioned in a blind spot where $\DaAvg|_{\fgw=3\fsat} \sim 1.31\times 10^{-3}$ (see Figure~\ref{fig:skymap}), the first harmonic becomes most dominant with $\DaAvg|_{\fgw=\fsat} \sim 0.5$.

The right panel of Figure~\ref{fig:CircEcc} presents results for the \texttt{GSAT0201} satellite with an eccentric orbit. We show two eccentricities: the actual value $e=0.162$, and a hypothetical case $e=0.4$ to emphasize the role of eccentricity. We see that eccentric orbits activate multiple harmonics. While the third harmonic remains dominant at lower eccentricity, higher harmonics become more significant as eccentricity increases. For instance, the fifth harmonic dominates at $e=0.4$, and the tenth at $e=0.6$. As the eccentricity increases, the number of contributing higher harmonics also grows, and these harmonics become significantly stronger at higher eccentricities. As a result, the signal from an inspiralling binary can occur in resonance multiple times throughout its evolution, delivering successive “kicks” to the satellite. These repeated interactions enhance the signal’s observability compared to a circular orbit.

\begin{figure*}
\centering
\includegraphics[width=0.97\textwidth ]{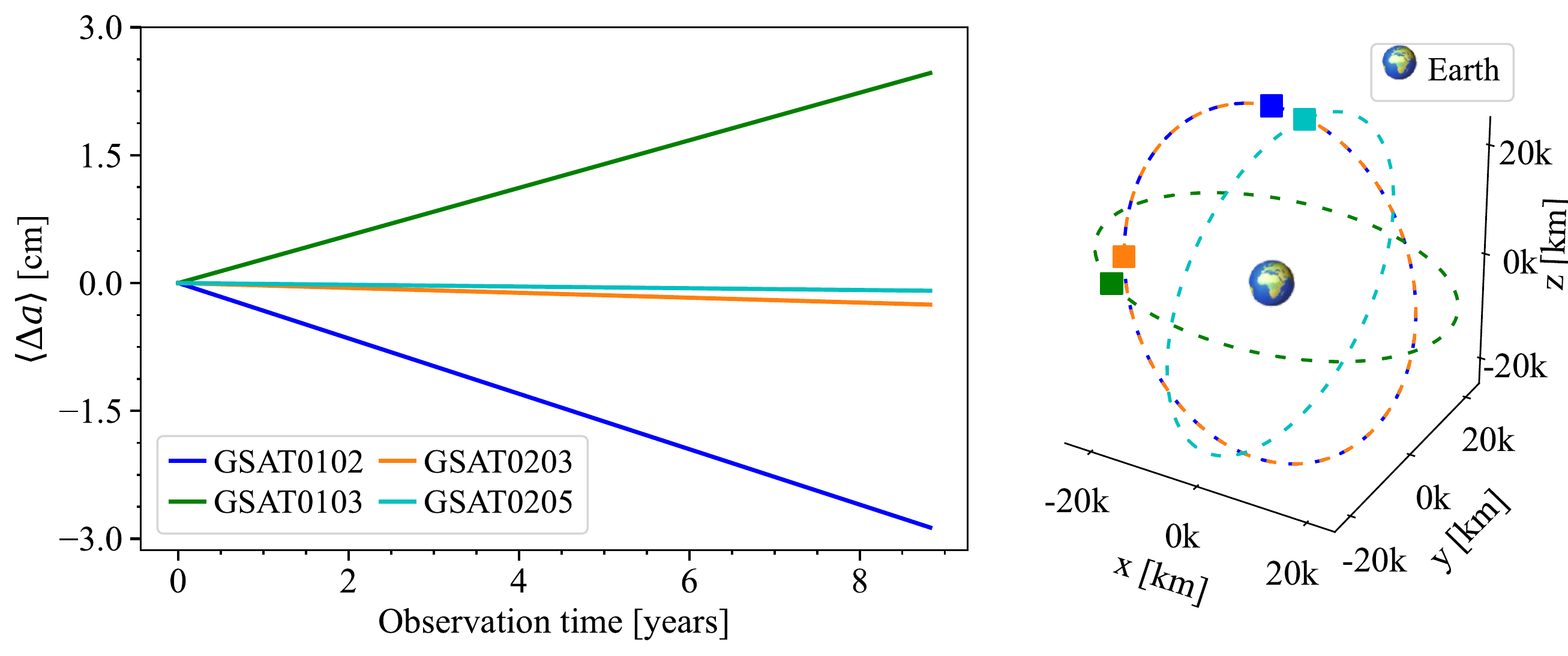}
\caption{Illustration of the evolution of the semi-major axis of a network of Galileo satellites under the influence of a resonance effect induced by a continuous GW signal at a frequency of $\fgw = 3 f_{\rm sat}$ and strain amplitude $h_0 \sim 10^{-14}$. The x-axis represents the observation time. For all the satellites, $\fsat \sim 1.97\times10^{-5}\:\rm{Hz}$.}
\label{fig:network}
\end{figure*}

Finally, we examine the GW-induced perturbation on a network of satellites. We consider four Galileo satellites: \texttt{GSAT0102}, \texttt{GSAT0103}, \texttt{GSAT0203}, and \texttt{GSAT0205}. All follow circular orbits with the same inclination angle $\iota$, but differ in orbital orientation parameter $\Omega$, except for \texttt{GSAT0102} and \texttt{GSAT0203}. These two have identical orbital elements but differ in angular position, as illustrated in the right panel of Figure~\ref{fig:network}. We evolve each orbit under the same monochromatic signal over a 10-year observation period. The left panel of Figure~\ref{fig:network} shows the secular drift in semi-major axis $\DaAvg$ over time. Depending on the phase relationship, the orbits may either expand or shrink, and the rate of change in $\DaAvg$ differs among satellites---even those with identical orbital parameters---due to differences in angular positioning.

\section{Conclusion}

We show that GNSS constellations can probe GWs in the frequency range of approximately $10^{-5}$ to $10^{-3}\:\rm{Hz}$, allowing us to target signals from binary black holes with masses from millions to billions of solar masses---well outside the LISA sensitivity band. Additionally, we can aim to detect stochastic GWs from supermassive black holes in galaxies and the early Universe.

We have developed a framework for calculating the deviation in the evolution of GNSS orbits induced by GW signals. The orbital deviation is amplified resonantly when a GW frequency is an integer multiple of the binary's orbital frequency. For a circular orbit, resonances occur up to the third harmonic, while eccentric orbits can activate resonances at many higher harmonics. 


Satellites are always subject to several systematic effects that can reduce the observability of GW signals. These include non-gravitational perturbations like solar radiation pressure and thermal forces, as well as clock noise, orbit determination errors, and signal propagation delays. Such effects can mask GW-induced signatures, lowering detection sensitivity unless carefully modeled and mitigated. Since GW-induced orbital deviations are coherent across the constellation, a satellite network enables disentangling GWs from systematics.

\section*{Acknowledgments}
We thank A. Hees and D. Blas for valuable discussions related to this project. S.R. acknowledges support from the Fonds de la Recherche Scientifique – FNRS (Belgium).

\section*{References}
\bibliography{moriond}


\end{document}